\documentclass{ifacconf}
\usepackage{graphicx}      % include this line if your document contains figures
\usepackage{natbib}        % required for bibliography
\usepackage{bm}
\usepackage{amssymb,amsmath}
\usepackage[utf8x]{inputenc}
\usepackage{comment}
\usepackage[export]{adjustbox}

\usepackage{xcolor}

\begin{document}
\begin{frontmatter}

\title{Active Sidestick Control Integration for Enhanced Aircraft Flight Envelope Protection}
%
%An Integrated Framework for the Development of Advanced Protections Features Coupled with Active Sidestick}
%\thanksref{footnoteinfo} 

%\footnote{This work has been submitted to IFAC for possible publication.}

\thanks[footnoteinfo]{This work has been submitted to IFAC for possible publication.}

\author[First]{Çağrı Ege Altunkaya, Fatih Erol},\footnotemark[1]
\author[First]{Akın Çatak},\footnotemark[2]
\author[Third]{Volkan Mert},\footnotemark[3]
\author[Second]{Pierluigi Capone},\footnotemark[4]
\author[Third]{Şükrü Akif Ertürk},\footnotemark[5]
\author[First]{Emre Koyuncu},\footnotemark[6]

\address[First]{Istanbul Technical University, Aerospace Research Center, 
   \\Istanbul, Türkiye.}
\address[Second]{Zurich University of Applied Sciences, Centre for Aviation, \\Winterthur, Switzerland.}
\address[Third]{Turkish Aerospace, Air Vehicle Technologies Research Center, \\Istanbul, Türkiye.}

\begin{abstract}                % Abstract of 50--100 words
The design of Envelope and Pilot-Induced Oscillation (PIO) Protection Features, and Failure Cases detection and prevention using Active Control Sidestick (ACS) is a challenging task. While helping the pilot to respect the envelope limitations also in failure scenarios and, therefore, increasing mission effectiveness, these features may have a significant impact on the aircraft's agility. ACS characteristics are investigated in an integrated environment. A set of effective and flexible control laws based on Incremental Nonlinear Dynamic Inversion have been developed in a state-of-the-art aircraft simulation model and coupled with a two-ways communication with the selected ACS. The model can run in real-time in a fixed-based simulator composed of representative cockpit and out-of-the-window.
\end{abstract}
\begin{keyword}
Active Sidestick, Flight Envelope Protection, Pilot-Induced Oscillation Protection, Incremental Nonlinear Dynamic Inversion, Integrated Aircraft Simulation Model.
\end{keyword}
\end{frontmatter}
%===============================================================================
%\section*{Acronyms / Symbols}
% \begin{table}[ht]
%    \centering
%    \begin{tabular}{ll}
%        CAS & Command Augmentation System \\
%        CHR & Cooper Harper Rating \\
%        CLAWS & Control Laws \\
%        EP & Envelope Protection \\
%        FCS & Flight Control System \\
%        HQ & Handling Qualities \\
%        MTE & Mission Task Elements \\
%        PI & Proportional Integral Controller \\
%        SAS & Stability Augmentation System \\
%        SCAS & Stability and Command Augmentation System \\
%        TCL & Thrust Control Lever \\
%        UDP & User Datagram Protocol \\
%    \end{tabular}
%    \caption{Acronyms and Symbols}
%    \label{tab:1}
% \end{table}
%%%%%%%%%%%%%%%%%%%%%%%%%%%%%%%%%%%%%%%%%%%%%%%%%%%%%%%%%%%%%%%%%%%%%%%%%%%%%%%%%%%
\footnotetext[1]{1st author (altunkaya16@itu.edu.tr; erolfa@itu.edu.tr)}\footnotetext[2]{2nd author (catak15@itu.edu.tr)}\footnotetext[3]{3rd author (volkan.mert@tai.com.tr)}
%\color{red}\textbf{Akif will revise! Max 1.5 pages}\color{black}\\

\section{Introduction}
\label{section:Introduction}
High performance aircraft are often designed at the leading edge of technology where mission performance can be heavily impacted by deficiencies. Therefore, everything which is possible must be done to reduce pilot workload, especially in situations with rapidly changing flight conditions based on standards (\cite{DOD:04}).
\par In these scenarios the pilot might not be able to monitor the boundaries of the flight envelope (FE) in an effective way. This results in increased risks and disadvantage since the available flight performance may not be fully used (\cite{RTO:00}). Aircraft limitations can be monitored by Envelope Protection (EP also called Carefree Handling) functions resulting in increased mission success due to reduced workload, more aggressive maneuvering, and reduced risk. Examples for EP functions from the literature are (\cite{fep2}), (\cite{fep5}), (\cite{fep4}) and,(\cite{fep1}).
\par Careful design must be applied to prevent possible disadvantages of EP such as reduced agility (\cite{AG:97}). Therefore, an integrated approach to these features where the EP algorithm is coupled with an ACS (\cite{INC:07} and (\cite{INC:18}), able to change its characteristics during the mission, is highly recommended (\cite{Amn18}). Similar reasoning can be made for PIO detection and prevention mechanisms. A more integrated solution will allow to cope with possible limitations of the aircraft system in specific points of FE. Additionally, possible failure conditions affecting ACS should be considered and coped with using detection and reconfiguration approaches.
\par The motivation of this work is to understand the ACS characteristics and to develop advanced flight protection algorithms within these scope. Therefore, an integrated framework shall be set up for further demonstrations.
\par The paper is organized as follows. In Section~\ref{section:Problem Statement}, the motivation of the work and the problem statement is described. Section~\ref{section:Integrated Active Sidestick Control System} details the integrated framework setup such as simulation models and all the components of the simulation environment. The architecture of flight envelope protection is explained in Section~\ref{section:Flight Envelope Protection}. Preliminary test results and the initial outcomes are given in Section~\ref{section:Simulation Setup and Results}. Finally, the conclusions are given in Section~\ref{section:Conclusion}.
%%%%%%%%%%%%%%%%%%%%%%%%%%%%%%%%%%%%%%%%%%%%%%%%%%%%%%%%%%%%%%%%%%%%%%%%%%%%%%%%%%%v
\footnotetext[4]{4th author (capo@zhaw.ch)}\footnotetext[5]{5th author (sukruakif.erturk@tai.com.tr)}\footnotetext[6]{6th author (emre.koyuncu@itu.edu.tr)}

\section{Problem Statement}
\label{section:Problem Statement}

Many current high-performance aircraft are equipped with Active Control Sidesticks (ACS). However, the details of these applications are often not publicly accessible. The design of Envelope Protection (EP), Pilot-Induced Oscillation (PIO) protection, and failure case detection and prevention using ACS is a challenging task. These features help pilots adhere to envelope limitations even in failure scenarios, thereby increasing mission effectiveness. However, they can also significantly impact the aircraft's agility. A thorough understanding of these issues is crucial for designing current and future aircraft more effectively and efficiently.

To address these challenges, several key research questions must be answered: What integrated infrastructure is needed to develop and test EP and PIO prevention algorithms? Which existing EP and PIO prevention algorithms are the most effective? What are the advantages and disadvantages of ACS? What are the likely failure scenarios of an ACS, and how can they be mitigated?

The proposed research aims to address these questions through a focus on five main areas: i) building understanding of ACS features and mathematical models, ii) integrating flight envelope protection (FEP), PIO detection and prevention algorithms, and iii) demonstration. This paper details the development of an integrated framework designed to answer these research questions and presents preliminary test results.

The framework comprises several key elements: the aircraft simulation model, flight control laws, active control sidestick, visual system, and cockpit. The aircraft simulation model is designed to be representative of modern high-performance aircraft, capable of running in real-time, interfacing with all other components, supporting detailed linear and non-linear analyses, and collecting and saving all simulation data. Flight control laws must be representative of current protection algorithms as benchmarks and should support advanced protection algorithms. The active control sidestick should incorporate state-of-the-art ACS features and enable two-way communication with the simulation model.

The visual system is required to operate in real-time with latency suitable for task evaluation and to represent dedicated mission task elements accurately. The cockpit should implement a Hands on Throttle and Stick (HOTAS) system and support touch-screen instrument panels. By integrating these components, the framework aims to address the challenges in designing effective EP and PIO protection features, thus enhancing the safety and performance of high-performance aircraft.

This integrated approach ensures a comprehensive understanding of ACS characteristics and the development of advanced flight protection algorithms, ultimately contributing to more efficient and safer aircraft designs.

\section{Integrated Active Sidestick Control System}
\label{section:Integrated Active Sidestick Control System}

\subsection{Aircraft Model}

In this section, the backbone dynamic, which is a generic F-16 model, are briefly presented, and these equations establish the foundation for the nonlinear flight dynamics model. Furthermore, the aerodynamic and actuator modeling are introduced with relevant references.

\subsubsection{Equations of Motion}
\label{eqofMot}

The rigid non-linear body dynamics is given in a compact form in Eq.~\eqref{translationalDynamics} and Eq.~\eqref{rotationalDynamics}. 

\begin{equation}
\label{translationalDynamics}
    \bm{\dot{V}} = m^{-1}[\bm{F} -  \bm{\omega} \times m \bm{V}]
\end{equation}  

\begin{equation}
\label{rotationalDynamics}
    \bm{\dot{\omega}} = J^{-1}[\bm{M} -  \bm{\omega} \times J \bm{\omega}]
\end{equation}  
where $\bm{V} \in \mathbb{R}^{3 \times 1}$ is the body velocity vector, $\bm{\omega} \in \mathbb{R}^{3 \times 1}$ is the angular rate vector, $\bm{F} \in \mathbb{R}^{3 \times 1}$ is the total body force vector, $\bm{M} \in \mathbb{R}^{3 \times 1}$ is the total moment vector, $J \in \mathbb{R}^{3 \times 3}$ is the inertia tensor, and $m$ is the mass of the aircraft. Additionally, the rotational kinematics are given by Eq.~\eqref{rotationalKinematics}.

\begin{equation}
\label{rotationalKinematics}
    \bm{\dot{\Omega}} = 
\begin{bmatrix}
1 & \sin\phi\tan\theta & \cos\phi\tan\theta \\
0 & \cos\phi & -\sin\phi \\
0 & \sin\phi \sec\theta & \cos\phi \sec\theta
\end{bmatrix} \bm{\omega}
\end{equation}  
where $\bm{\Omega}$ denotes the Euler angles.

\subsubsection{Aerodynamics}
\label{aeroModel}

Aerodynamic modeling is constructed using the data provided in \cite{f16Data}, which consists of wind tunnel test data conducted up to 0.6 Mach. All the data provided are implemented into interpolation lookup tables. After interpolation, the force and moment coefficients are calculated according to the formulation given in the same document \cite{f16Data}, and dimensionalization is carried out using the necessary atmospheric and geometric properties.

\subsubsection{Actuators}
\label{actuatorModel}

The dynamics of actuators are treated as a first-order system with the time-constant, rate, and position saturation constraints, as specified in \cite{f16Data}. Each control surface has a time constant of $0.0495s$ (that gives circa $20$ rad/s), whereas rate saturations are $60^\circ/s$, $80^\circ/s$, and $120^\circ/s$ for the horizontal tail, ailerons, and rudder respectively. Moreover, the position saturation are $\pm 25^\circ$, $\pm 21.5^\circ$, and $\pm 30^\circ$ for the horizontal tail, aileron, and rudder, respectively.

\subsection{Flight Control}
The control augmentation system consists of a single-loop angular rate control law using the incremental nonlinear dynamic inversion (INDI). The derivation of the control law is quite straightforward due to the control-affine form of Euler's equations of motion, given by Eq.~\eqref{rotationalDynamics2} in a decomposed form.

\begin{equation}
\label{rotationalDynamics2}
    \bm{\dot{\omega}} 
    = 
    -J^{-1} (\bm{\omega} \times J \bm{\omega})
    + 
    \underbrace{J^{-1}\bar{q}_\infty S 
    \begin{bmatrix}
      b &  &  \\
      & \bar{c} &   \\
      &   & b \\
    \end{bmatrix} \Phi}_{\substack{\bm{g}(\bm{x})}} 
    \underbrace{\bm{\delta}}_{\substack{\bm{u}}} 
\end{equation} 
where $\Bar{q}_\infty$, $S$, $b$, and $\Bar{c}$ are the dynamic pressure, wing area, wing span, and mean aerodynamic chord, respectively. Additionally, $\Phi \in \mathbb{R}^{3 \times n}$ represents the control effectivity matrix, which consists of the moment coefficient derivatives with respect to control surface deflections at the current instant, with $n$ denoting the number of control surfaces. The INDI control law for the control of the angular rates is derived in the form given in Eq.~\eqref{INDIlaw} (for further details, please see \cite{INDIlaw}),
\begin{equation}
\label{INDIlaw}
    \bm{u} = \bm{g}(\bm{x}_0)^{-1}[\bm{\Dot{\omega}}_c - \bm{\Dot{\omega}}_0] + \bm{u}_0 
\end{equation} 
where the subscript $"0"$ denotes the current state and $\bm{\Dot{\omega}}_c \in \mathbb{R}^3$ is the virtual input to be designed, the finalized form of the control law is given in Eq.~\eqref{INDIlaw2}. 

\begin{equation}
\label{INDIlaw2}
    \bm{\delta} = \Bigg\{J^{-1}\bar{q}_\infty S 
    \begin{bmatrix}
      b &  &  \\
      & \bar{c} &   \\
      &   & b \\
    \end{bmatrix} \Phi \Bigg\}^{-1}[\bm{\Dot{\omega}}_c - \bm{\Dot{\omega}}_0] + \bm{\delta}_0 
\end{equation} 
where $\bm{\delta} \in \mathbb{R}^3$ signifies the control surface deflections, i.e. horizontal tail, aileron, and rudder, respectively. Moreover, since there are also three control surfaces ($n = 3$), the control effectivity matrix is a square matrix, that is, it is invertible unless it is rank deficient. Furthermore, the virtual input $\bm{\Dot{\omega}}_c$ is provided by Eq.~\eqref{eq:fca13}.

\begin{equation} \label{eq:fca13}
\bm{\Dot{\omega}}_c
= 
\begin{bmatrix}
K_p & & \\
 & K_q & \\
 & & K_r \\
\end{bmatrix} 
\begin{bmatrix}
p_{c} - p \\
q_{c} - q \\
r_{c} - r \\
\end{bmatrix} 
\end{equation}
where $K_p$, $K_q$, and $K_r$ represent the gains for the roll, pitch, and yaw channels, respectively. The expressions presented provide the required generation of control surface deflections with respect to the pilot inputs of $p_c$, $q_c$, and $r_c$. 

Consequently, the aforementioned relations complete the flight control law for the angular rates. The constructed flight control architecture is straightforward and flexible, allowing for seamless integration with other components and subsystems, e.g. it can accommodate various flight envelope protection algorithms and active sidestick features, facilitating the realization of a comprehensive pilot-in-the-loop environment.

\subsection{Active Sidestick Control}
This section explains the characterization of the active side-stick control system, which plays a crucial role in improving pilot-aircraft interactions and enabling flight envelope protection features.

STIRLING DYNAMICS\textsuperscript{\textregistered} Compact Stick Active Inceptor is a state-of-the-art low-force development ACS and uses a mass-spring-damper model with adjustable parameters.
The Inceptor Control Module (ICM) is an integrated microcomputer and brushless DC motor drive designed to be used as a common control module for a variety of compatible active control inceptors. The ICM runs the control law for the force feedback system and interfaces with the host computer in the simulator system.
\par The ACS grip has four switches: S1 (left), S2 (right), S3 (trim), and S4 (trigger). The default power-up mode is Disabled Mode. After the Initial Built-In Test (IBIT), the system zeroes force measurements. In Enabled Mode, the system can be commanded to provide artificial force feedback. In Jammed Mode, the motors can be commanded to lock the motion of the stick. Message transmission via UDP on Ethernet occurs only with command changes at 0-200 Hz for all axes. ACS features are sent through ID1-12s \& 50 and read over ID20-24s as identification number, (ID) messages.

\begin{figure}[ht]
    \centering
    \includegraphics[width=1\linewidth]{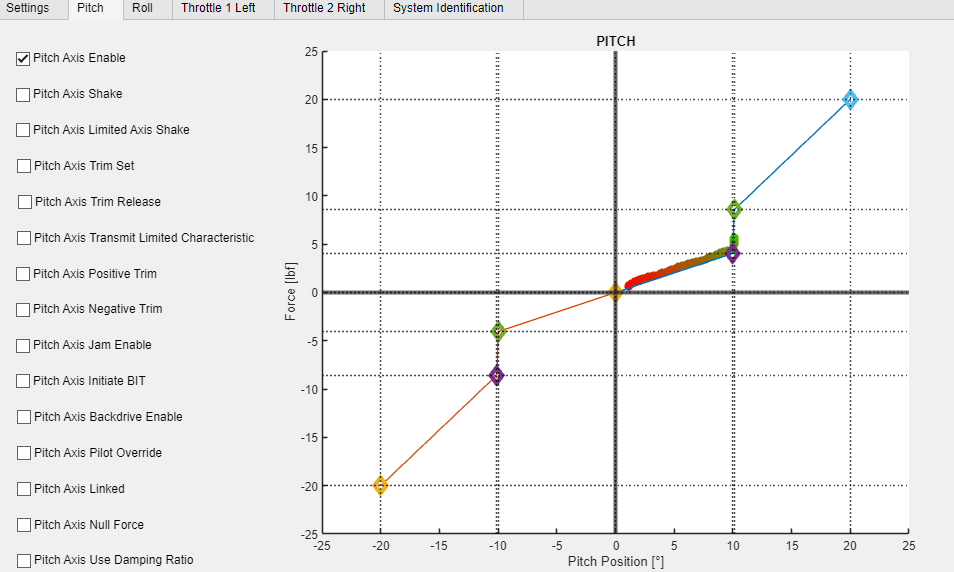}
    \caption{Force Feel Characteristics Curve and Enable Flags}
    \label{fig:FFC and Enable Flags}
\end{figure}
\par Control message (ID2) manages features including trim, shaker, and damping coefficients. The Force Feel Characteristics Curve (FFC) requires increasing force/position values. Characteristic Control (ID5) sets up the fade time when transitioning between different FFC. Trim Control (ID6) adjusts axis trim positions. Shaker Control (ID7) specifies shaker amplitude and frequency. Damping Control (ID8) sets the damping ratio (0.2-6). Cueing Force Control (ID9) adjusts cueing force characteristics. Inertia Control (ID10) defines perceived inertia (0.01-10 lbf/in²). Two ACS units can be linked for two-pilots configuration.
\par Rotary Status Message (ID20) reads axes status at set rates, including mode and grip switch states. Limited Characteristic Message (ID22), 'limited,' restricts control characteristics due to motor capabilities.  IP Address Change (ID50) modifies network addresses. 
\par Both pitch and roll axes have 24° position and a 27 lbf force limits, as this is a Low Force ACS. An example pitch axis configuration is shown in Fig.\ref{fig:FFC and Enable Flags}.
\par The system's dynamics are modeled through a second-order transfer function with parameters including FFC, Damping Ratio, Moment of Inertia (MOI), and friction. The governing equation of motion is
\begin{equation}
\label{eom}
m\ddot{\theta} + c\dot{\theta} = F_{char} + F_{friction} + F_{grip}
\end{equation}
where \(m\) denotes the MOI adjusted by 0.01, and \(c\) is calculated from the damping ratio \(\zeta\) and force gradient \(k\). Assuming no friction and fixed \(k\), the model simulates the sidestick's response to grip force \( F_{grip}(s) \).

\begin{equation}
\label{pitchangletheta}
\mathbf{\theta(s)} = \frac{1}{m s^{2} + (2\zeta \sqrt{m k}) s + k} \mathbf{F_{grip}(s)}
\end{equation}
where \(m = 0.6\), \(\zeta = 0.35\), \(k = 2\) and \(F_{grip} = 11 lbf\).

% Fatihin Kısım
\par
\begin{figure}[ht]
    \centering
    \includegraphics[width=0.75\linewidth]{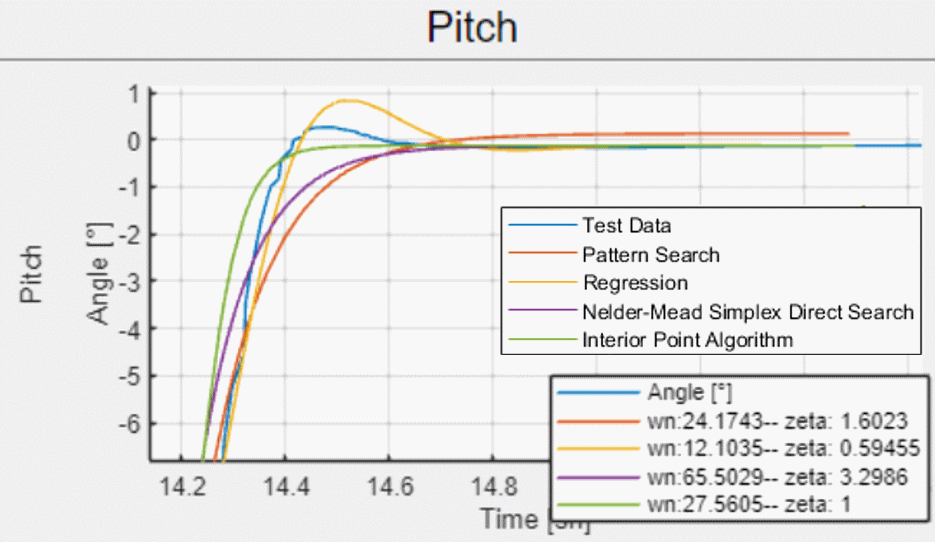}
    \caption{The ACS Parameter Identification}
    \label{fig:The ACS Parameter Identification}
\end{figure}

 Although the ACS is similar to a mass-spring-damper system, it is a combined system with complex electrical and magnetic features. A set of different fitting methods have been used to investigate this highly nonlinear behaviour. First, the stick was moved to the deflection of 10° and then released. The recorded data was fitted with a 2nd order transfer function with 4 different approaches. Noise and system nonlinearity cause discrepancies between optimized and calculated damping and frequency values, indicating practical deviations from the theoretical model as illustrated in Fig. \ref{fig:The ACS Parameter Identification}.

\section{Flight Envelope Protection}
\label{section:Flight Envelope Protection}

To integrate with the ASC, a couple of conventional FEP algorithms are developed and implemented. The protection algorithms for angular rates, angle of attack, and bank angle are developed based on a common and simple rationale: applying a counter-action as a restorative measure to prevent FEP violations.

\subsubsection{Pitch, Roll, and Yaw Rate Protection}

Since the commanded pilot inputs are also the angular rates, it is simply possible to restrict the pilot commands with a dynamic saturation, which is fed by the scheduled flight envelope database, to prevent a possible excess of the flight envelope. Subsequent to subjecting the pilot commands of $\omega_p$ to the maximum and minimum allowable values of the flight envelope, the resulting commanded angular rates of $\omega_c$ are then applied to the control augmentation system.

\subsubsection{Angle of Attack and Load Factor Protection}

\begin{figure}[hbt!]
\centering
\includegraphics[width=3.4in]{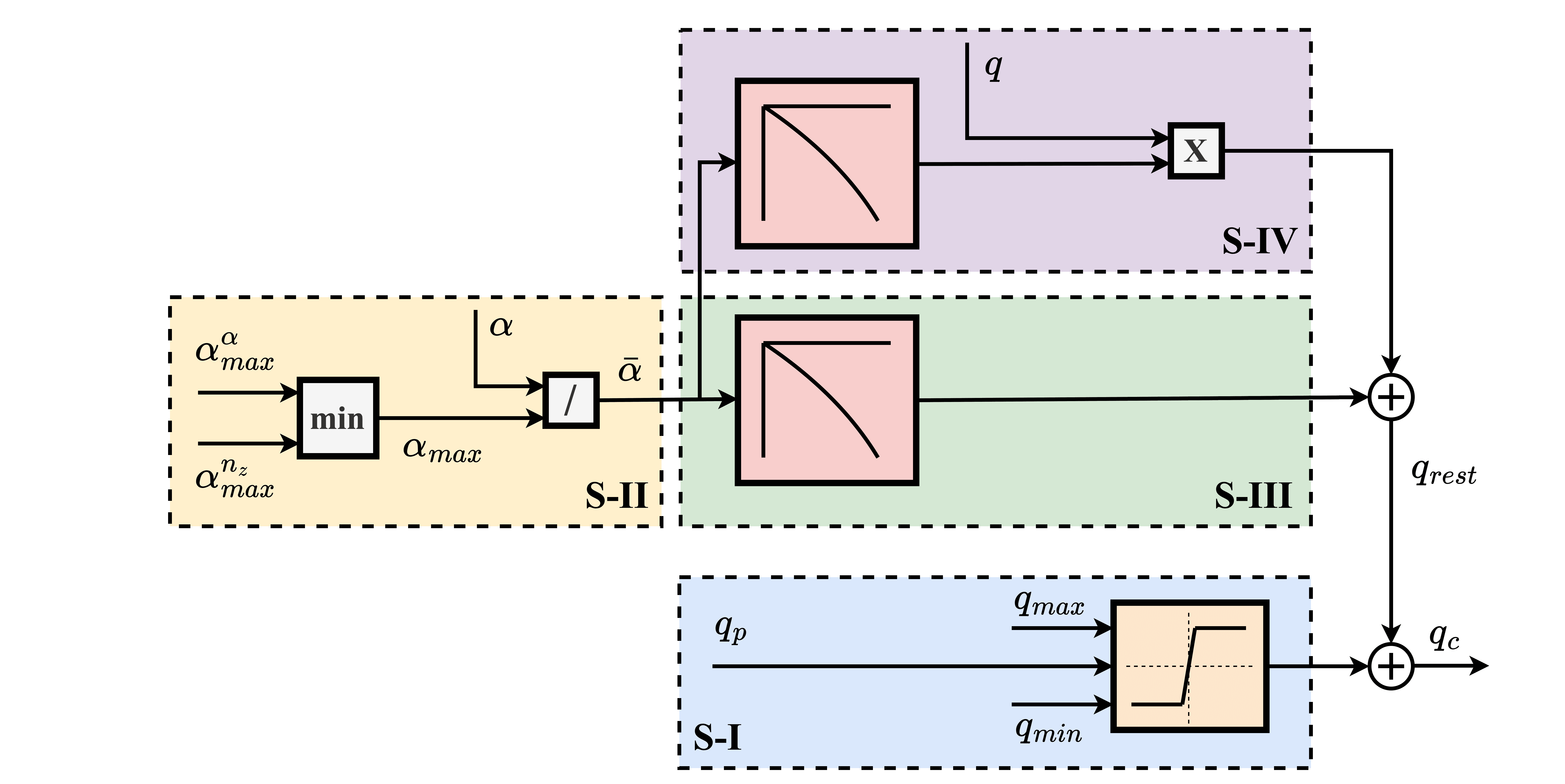}
\caption{The angle of attack and load factor protection scheme: four main components, S-I/II/III/IV.}
\label{fig:longProt}
\end{figure}

The violation of the angle of attack and load factor boundaries can be avoided by applying a 'restorative' pitch rate command. Prior to reaching the predetermined limits of the angle of attack and load factor, the pilot commands are faded out in order to prevent the aircraft from entering into an upset condition. Furthermore, both the angle of attack and load factor protection should be considered holistically to generate a single and unified restorative pitch action; therefore, the load factor boundary is also converted to the angle of attack as given by Eq.~\eqref{loadFactor2Alpha}.

\begin{equation}
\label{loadFactor2Alpha}
    \alpha_{max}^{n_z} = \dfrac{W n_{z_{max}}}{\Bar{q}_\infty S C_{z_\alpha}}
\end{equation}
where $\alpha_{max}^{n_z}$ denotes the angle of attack equivalent to the maximum allowable load factor $n_{z_{max}}$, $W$ is the weight of the aircraft, $C_{z_\alpha}$ is the force coefficient in z-direction derivative with the angle of attack. Afterward, the minimum of the main angle of attack limits, $\alpha_{max}^{\alpha}$, and equivalent angle of attack limit, $\alpha_{max}^{n_z}$, is used to generate the restorative action. The general framework is depicted in Fig.~\ref{fig:longProt}

S-I is the pitch rate protection component, which has been discussed in the preceding section. S-II is the logic of the selection of the restrictive angle of attack; furthermore, the S-III and IV consist of designed restorative actions. To ensure the generality of the generated action in terms of magnitude, normalization is performed in S-II. The normalized angle of attack indicator, $\Bar{\alpha}$, reflects the current $\alpha$ as the percentage of $\alpha_{max}$. This normalization enables the generation of a restorative pitch rate based on the proximity to the limit. Consequently, in S-IV, a damping component is incorporated to reduce the magnitude of oscillation during protection, leveraging the current pitch rate, $q$. Note that this structure is highly adaptable for design, and a reverse rationale can be applied to protect against violations of the minimum angle of attack and load factor limits as well.

\subsubsection{Bank Angle Protection}

\begin{figure}[hbt!]
\centering
\includegraphics[width=3in]{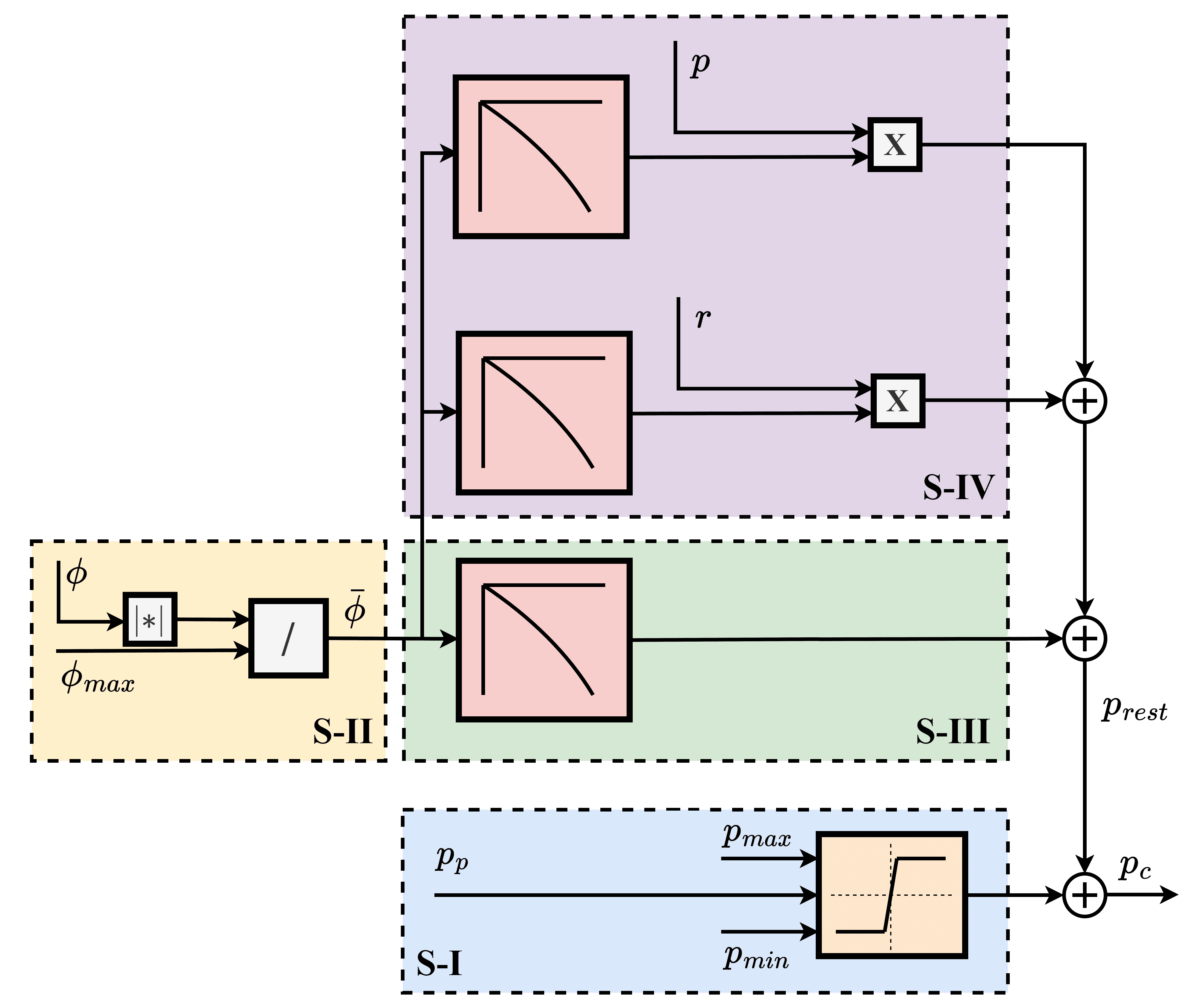}
\caption{The bank angle protection scheme: four main components, S-I/II/III/IV.}
\label{fig:bankProt}
\end{figure}

The bank angle protection is established with the same philosophy as the previous section. Simply, a restorative roll action is applied to not exceed the predetermined maximum allowable bank angle limit. The general framework is illustrated in Fig.~\ref{fig:bankProt}.

S-I is the roll rate protection component, while S-II is the normalization component considering the absolute of the current bank angle, $\phi$. Moreover, the S-III and IV, again, consists of designed restorative actions. In S-III, a counter-action roll rate is generated based on the current bank angle as the percentage of the allowable limit. In addition, in S-IV, two distinct damping components are incorporated to fade the magnitude of oscillation out during protection, using both the current roll rate, $p$, and the yaw rate, $r$. However, the S-IV component not only addresses the roll motion-induced bank angle build-up but also accounts for the possibility of yaw motion-induced bank angle build-up. Therefore, measures are taken to protect the bank angle envelope in both scenarios.

\section{Simulation Setup and Results}
\label{section:Simulation Setup and Results}

\par
\begin{figure}[ht]
    \centering
    \includegraphics[width=1\linewidth]{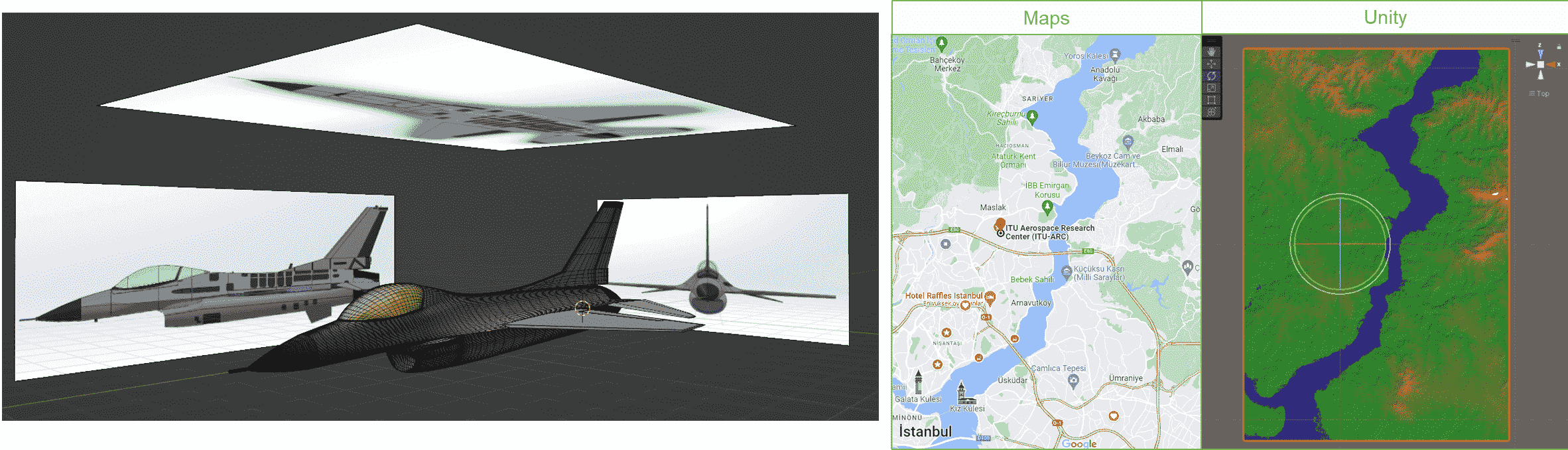}
    \caption{F-16 3D CAD Model used in the Simulation Environment and Terrain Map Model in UNITY\textsuperscript{\textregistered}}
    \label{fig:F16 3D Model and Unity}
\end{figure}

 The Visualization System runs in UNITY\textsuperscript{\textregistered} which is a cross-platform game engine developed by Unity Technologies\textsuperscript{\texttrademark} that is primarily used to develop video games and simulations for computers, consoles, and mobile devices. The F-16 3D model has been drawn in BLENDER\textsuperscript{\textregistered} in Fig. \ref{fig:F16 3D Model and Unity}. All control surfaces have been individually separated and rigged with bones to enable independent rotation.

%  \begin{figure}[ht]
%      \centering
%      \includegraphics[width=0.75\linewidth]{Visualization Files/4.png}
%      \caption{F16 3D Model: Mesh Count}
%      \label{fig:F16 3D Model: Mesh Count}
%  \end{figure}

%  \begin{figure}[ht]
%      \centering
%      \includegraphics[width=0.75\linewidth]{Visualization Files/2.png}
%      \caption{Aircraft Model in Blender and Unity}
%      \label{fig:Aircraft Model in Blender and Unity}
%  \end{figure}

% \par 3D model are transferred from Blender to Unity as illustrated in Fig. \ref{fig:Aircraft Model in Blender and Unity}.

\par Istanbul is selected for terrain map and Istanbul Technical University is selected for reference axes. Marmara region digital elevation data is used for recreating terrain in UNITY\textsuperscript{\textregistered} as illustrated in Fig. \ref{fig:F16 3D Model and Unity}.
Terrain information from Digital Terrain Elevation Data (DTED) contains only point clouds. Triangle meshes are created
between these point clouds for visualization.
%\begin{figure}[ht]
%    \centering
%    \includegraphics[width=0.75\linewidth]{Visualization Files/7.png}
%    \caption{Terrain Generation with Triangle Meshes in Unity}
%    \label{fig:Terrain Generation with Triangle Meshes}
%\end{figure}

%\par Terrain information from DTED contains only point clouds. Triangle meshes created
%between these point clouds for visualization as illustrated in
%Figure \ref{fig:Terrain Generation with Triangle Meshes}.

% \begin{figure}[ht]
%     \centering
%     \includegraphics[width=0.75\linewidth]{Visualization Files/9.png}
%     \caption{Terrain Top View and Terrain Occlusion}
%     \label{fig:Terrain Top View and Terrain Occlusion}
% \end{figure}

%\par To optimize performance, terrain occlusion has been incorporated, ensuring that only
%the closest terrain components are activated during simulation as illustrated in Fig.
%\ref{fig:Terrain Top View and Terrain Occlusion}.

%\begin{figure}[ht]
%    \centering
%\includegraphics[width=0.75\linewidth]{Visualization Files/3.png}
%    \caption{Overall View of F-16 Visual Interface}
%    \label{fig:Overall View of F16 Visual Interface}
%\end{figure}

%\par The Final visual interface can be seen as illustrated in Fig. \ref{fig:Overall View of F16 Visual Interface}.

% Fatihin Kısım Bitti.

\subsection{Aircraft Simulator Setup}
\begin{figure}[ht]
    \centering
    \includegraphics[width=0.75\linewidth]{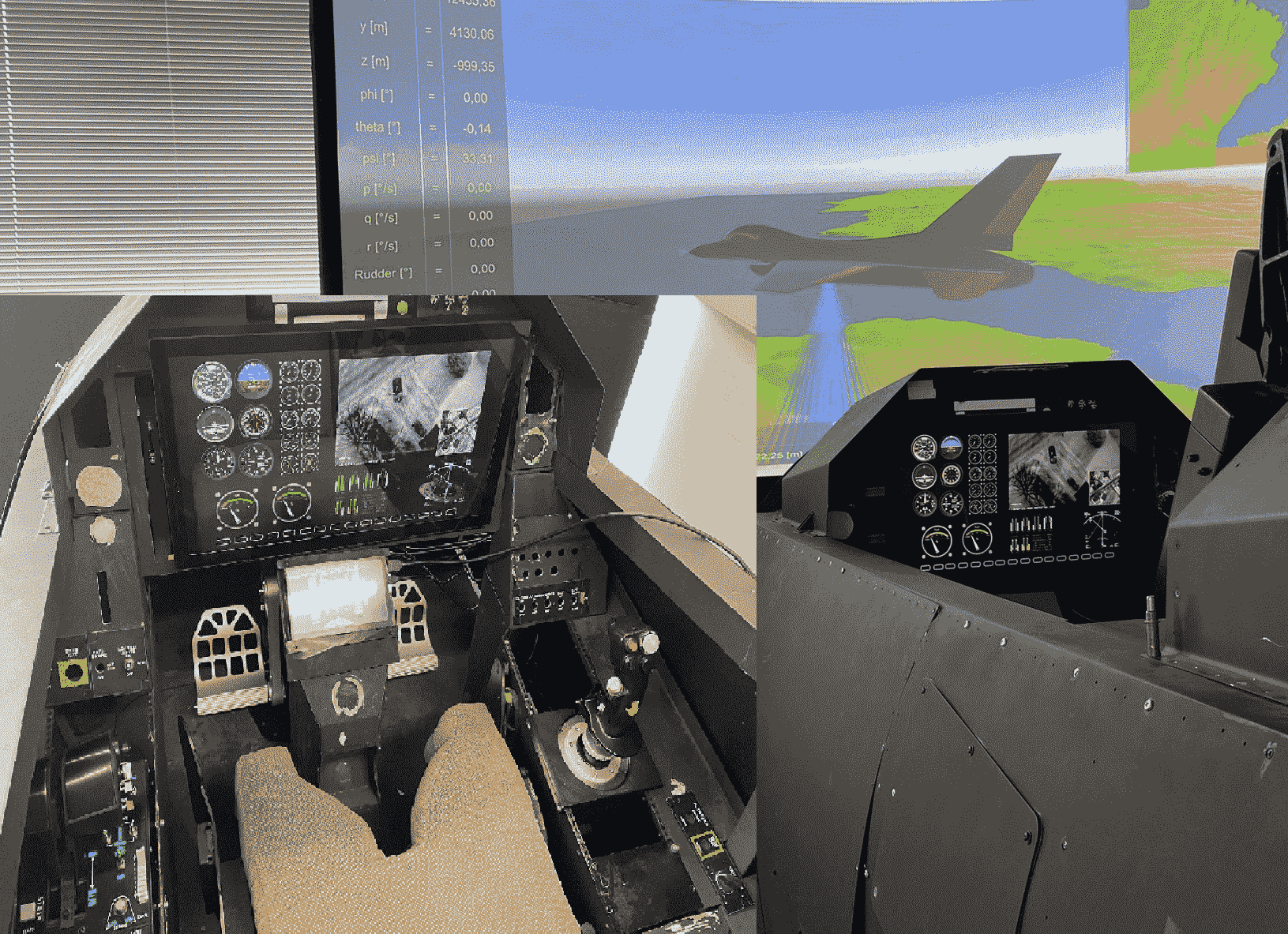}
    \caption{General view of Simulator Setup}
    \label{fig:Simulator}
\end{figure}

The aircraft cockpit shell is equipped with an 8x20 touch screen large area display (LAD) as instrument panel, a tablet used as control panel, and a THRUSTMASTER\textsuperscript{\textregistered} Hotas Wartog Hands-on-Throttle-and-Stick (HOTAS) system, which includes a sidestick, a dual rotary throttle, and two pedals. The general view of the integrated simulation environment is shown in Fig.~\ref{fig:Simulator}.
%\par Upon completion of on-the-desk sidestick communication studies, the HOTAS stick temporarily installed in the simulator, will be permanently replaced by the STIRLING DYNAMICS\textsuperscript{\textregistered} is a low force state-of-the art development ACS.
\subsection{ACS Component Interfaces}
% 1 column space allocation for the authors
%\par Prior to the development of MATLAB\textsuperscript{\textregistered} scripts, the signals transmitted by
\par
\begin{figure}[ht]
    \centering
    \includegraphics[width=0.9\linewidth]{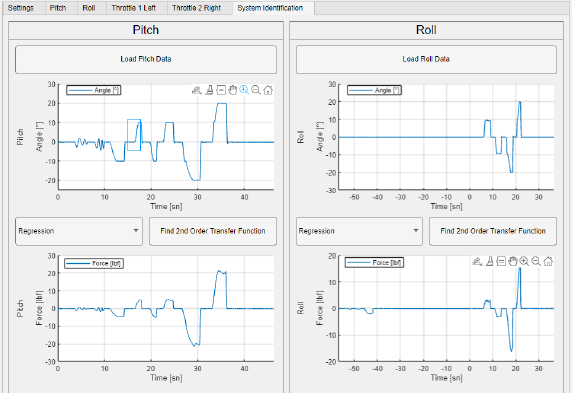}
    \caption{Example Signal Acquisition for System Identification  (pitch \& roll sweeps)}
    \label{fig:System Identification}
\end{figure}

 For system identification purpose of ACS, input and output signals were decoded, and tested using a network packet analyzer software that captures, aggregates, and reads data packets between local and target devices, such as Ethernet networks or recorded files as in Fig.~\ref{fig:System Identification}. The preliminary phase also involved comprehensive diagnostics on the Low Force Inceptor, utilizing STIRLING DYNAMICS\textsuperscript{\textregistered}’ IBIT software to confirm system integrity and operational safety. 
%\par Subsequently, UDP socket reading was conducted with MATLAB\textsuperscript{\textregistered}'s 'udpExplorer' application.
\par
\begin{figure}[ht]
    \centering
    \includegraphics[width=0.75\linewidth]{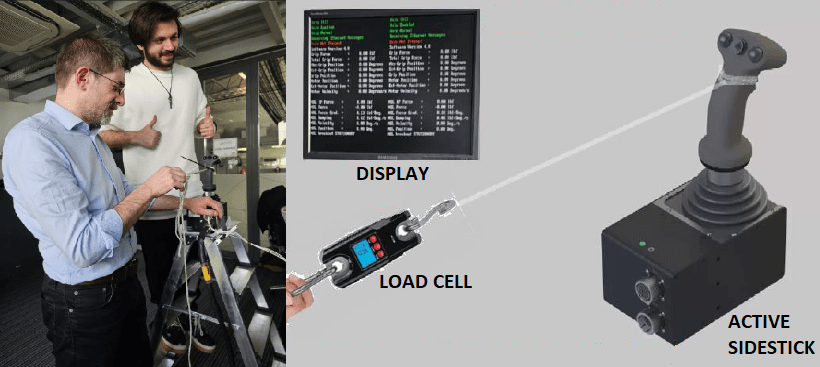}
    \caption{Calibration Tests with Load Cell Setup}
    \label{fig:Calibration Test}
\end{figure}

 A key aspect of the testing was the load calibration, which achieved a precision level of \%2 as shown in Fig.~\ref{fig:Calibration Test}. This was conducted using a load cell positioned at the detent button, facilitating accurate force measurements across control positions. Additionally, the control stick’s response was evaluated from a constrained negative, down-leftmost position to a positive, upright-most position, emphasizing the robustness of the high force feedback mechanism.
\par Ad hoc tests were instrumental in pinpointing uncertainties, thereby establishing definitive operational parameters, including minimum and maximum force limits and positional thresholds.
\par The aforementioned tests not only demonstrated the reliability of the ACS control but also revealed the identification of the axes through the binary analysis of the message header bits. This allowed for the verification of any bits that may have been overlooked between the pitch and roll axis readings. The observed discrepancies led to the implementation of alternative methodologies in subsequent development stages.
\par Latency was evaluated by monitoring UDP messages with scripts, which indicated a comparative decrease in efficiency against SIMULINK\textsuperscript{\textregistered} Real-Time Blocks. Thus, it is observed that the SIMULINK\textsuperscript{\textregistered} Real-Time Blocks are working faster. The integration of 'Matlab Functions' within SIMULINK\textsuperscript{\textregistered}  was also used to improve performance.

\par
\begin{figure}[ht]
    \centering
    \includegraphics[width=1\linewidth]{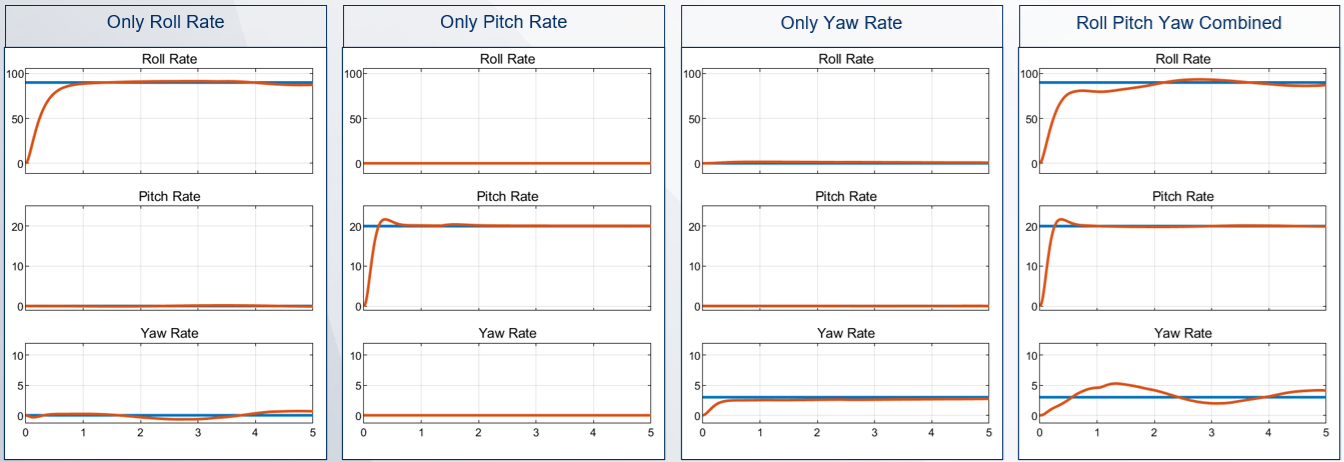}
    \caption{Controller Outputs Integrated with Active Sidestick Control}
    \label{fig:controller_performance}
\end{figure}

\begin{figure}[ht]
    \includegraphics[width=1.10\linewidth, right]{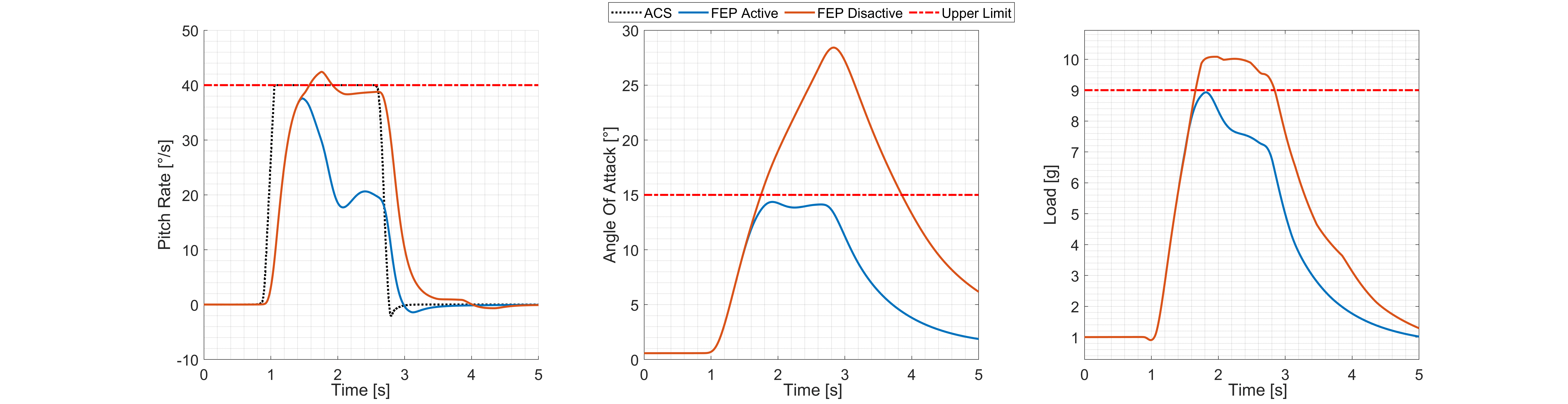}
    \caption{Envelope Protection in Longitudinal Axis}
    \label{fig:env_prot_long}
\end{figure}

 The commanded angular rates and their responses with respect to time are shown in Fig.  ~\ref{fig:controller_performance}. The performance of the flight control law is satisfactory in both decoupled and coupled pilot inputs. The affect of envelope annihilation and the effect of flight envelope protection structure is given in Fig. \ref{fig:env_prot_long} for longitudinal axis.

\par Although the performance of the INDI controller is satisfactory it may lead to envelope exceedance if certain pitch rate command is demanded from pilot. The red lines in the Fig. \ref{fig:env_prot_long} shows the limits in pitch, angle of attack and load factor. Even the pitch command is within the limits, it may result a in angle of attack or load factor exceedance. The results show that the proposed envelope protection algorithm prevents the exceedance before it happens.

\begin{figure}[ht]
\centering
\includegraphics[width=0.75\linewidth]{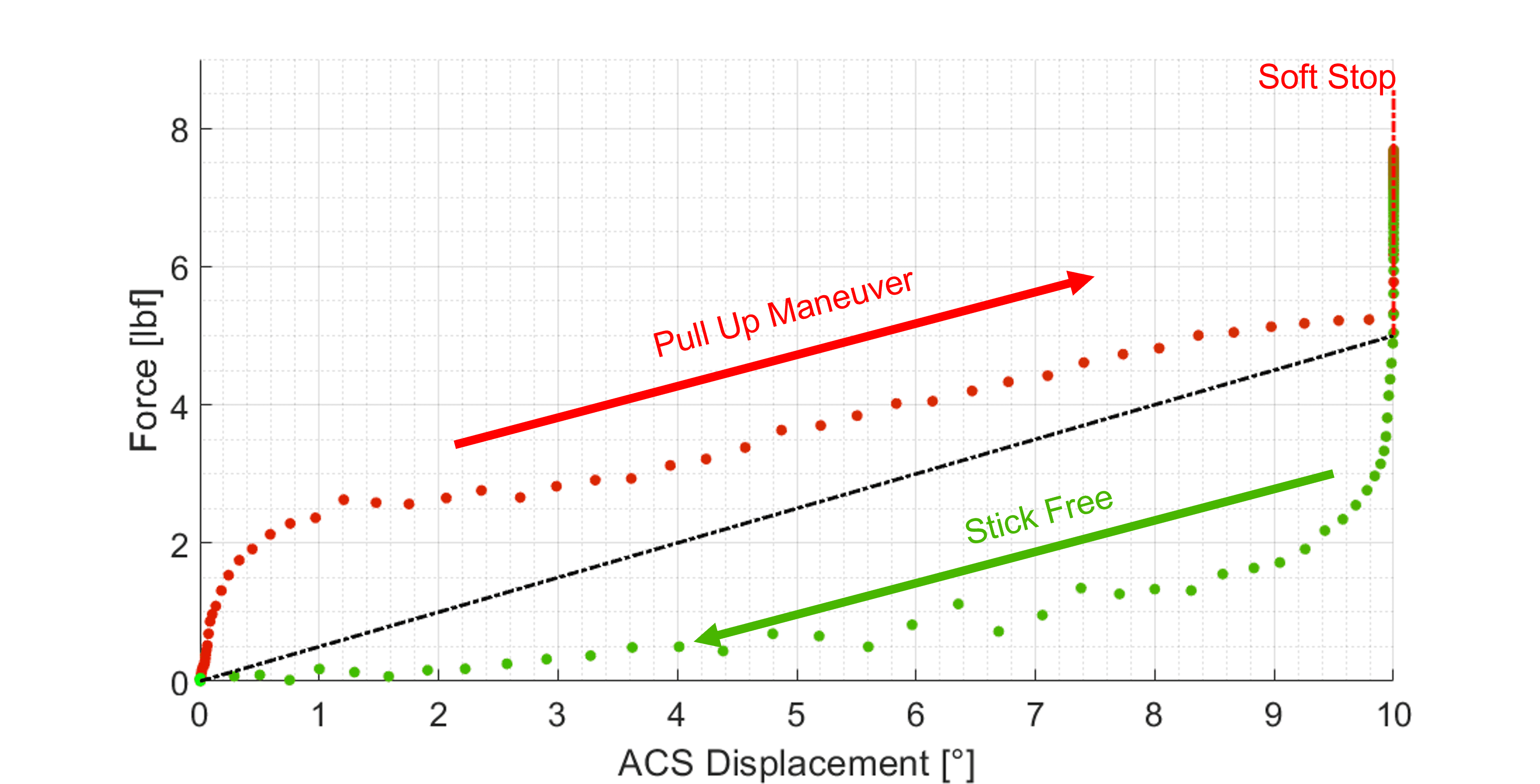}
    \caption{ACS Displacement-Force Curve}
    \label{fig:Force Displacement}
\end{figure}

The corresponding response of the active sidestick for the results shown in Fig.\ref{fig:env_prot_long} is given in Fig.\ref{fig:Force Displacement} to give the force displacement curve.

\section{Conclusion \& Outlook}
\label{section:Conclusion}

This work looked on how ACS technologies can be effectively integrated in a modern Flight Control System. The focus has been on FEP functionalities and on PIO protection. Another important aspect has been to verify how ACS works with pilot inputs. This is important for making aircraft safer and more efficient. The limitations of current ACS can be identified and the corresponding pitfalls in the development of effective integrated solutions should, therefore, be avoided. \par
An integrated environment that takes into accounts all the different aspects and components of the problem has been built and is currently under testing. In particular, a state-of-the-art simulation model of a high-performance aircraft has been coupled with effective and versatile control laws that will allow to develop "standard" and modern/advanced EP and PIO protection mechanisms. The model has been interfaced with 2-ways connections with a configurable ACS and also with a representative cockpit and out-of-windows.
\par
Preliminary testing give encouraging results and confidence in the capability of the integrated framework to face the challenges of designing more easier to fly, better controllable and, ultimately, more safe aircrafts.

% 1 column space allocation for the authors

%\begin{ack}
%Place acknowledgments here.
%\end{ack}

\bibliography{ifacconf}             % bib file to produce the bibliography
                                                     % with bibtex (preferred)
                                                   
%\begin{thebibliography}{xx}  % you can also add the bibliography by hand

%\bibitem[Able(1956)]{Abl:56}
%B.C. Able.
%\newblock Nucleic acid content of microscope.
%\newblock \emph{Nature}, 135:\penalty0 7--9, 1956.

%\bibitem[Able et~al.(1954)Able, Tagg, and Rush]{AbTaRu:54}
%B.C. Able, R.A. Tagg, and M.~Rush.
%\newblock Enzyme-catalyzed cellular transanimations.
%\newblock In A.F. Round, editor, \emph{Advances in Enzymology}, volume~2, pages
%  125--247. Academic Press, New York, 3rd edition, 1954.

%\bibitem[Keohane(1958)]{Keo:58}
%\newblock \emph{Power and Interdependence: World Politics in Transitions}.
%\newblock Little, Brown \& Co., Boston, 1958.

%\bibitem[Powers(1985)]{Pow:85}
%\newblock Is there a way out?
%\newblock \emph{Harpers}, pages 35--47, June 1985.

%\bibitem[Soukhanov(1992)]{Heritage:92}
%\newblock \emph{{The American Heritage. Dictionary of the American Language}}.
%\newblock Houghton Mifflin Company, 1992.

%\end{thebibliography}

%\appendix
%\section{A summary of Latin grammar}    % Each appendix must have a short title.
%\section{Some Latin vocabulary}              % Sections and subsections are supported  
                                                                         % in the appendices.
\end{document}